\documentclass[10pt]{article}
\usepackage[latin1]{inputenc}
\usepackage{multicol}
\usepackage{amsmath,amssymb}
\usepackage[english]{babel}
\usepackage{layout}
\usepackage{enumerate}
\usepackage{hyperref}
\usepackage{color,graphicx,graphics}
\definecolor{green2}{rgb}{0,0.83,0}
\hypersetup{
            colorlinks={true},
            filecolor={0.19,0.27,0.69},
            urlcolor={blue},
            linkcolor={red},
            citecolor={green2},
            }
\newcommand\email[1]{{\tt\href{mailto:#1}{#1}}}
\numberwithin{equation}{section}
\begin{document}
\begin{center}\Large{\textsf{\textbf{Geodesic Deviation Equation in $\boldsymbol{f(R)}$ Gravity}}}\\
\end{center}
\begin{center}
Alejandro Guarnizo, \footnote{\email{aguarnizot@unal.edu.co}} Leonardo Casta\~neda \footnote{\email{lcastanedac@unal.edu.co}} \&  Juan M. Tejeiro \footnote{\email{jmtejeiros@unal.edu.co}}
\end{center}
\begin{center}
\textit{Grupo de Gravitaci\'on y Cosmolog\'ia, Observatorio Astron\'omico Nacional,\\ Universidad Nacional de Colombia}\\
\textit{Bogot\'a-Colombia}
\end{center}
\date{}
\begin{abstract}
\noindent In this paper we study the Geodesic Deviation Equation (GDE) in metric $f(R)$ gravity. We start giving a 
brief introduction of the GDE in General Relativity  in the case of the standard cosmology. Next we generalize the GDE 
for metric $f(R)$ gravity using again the FLRW metric. A generalization of the Mattig relation is also obtained. Finally we 
give and equivalent expression to the Dyer-Roeder equation in General Relativity in the context of $f(R)$ gravity.
Accepted for Publication in \textit{\textbf{General Relativity and Gravitation}}\\\\
DOI: {\color{red}10.1007/s10714-011-1194-6}
\end{abstract}
\section{Introduction}
In the context of General Relativity (GR), the curvature and geometry of the space-time play a fundamental role: instead of forces acting in the Newtonian theory
we have the space-time curved by the matter fields \cite{Misner}.  The space-time could be represented by a pair $(\mathcal{M},\mathbf{g})$ with  $\mathcal{M}$  
a 4-dimensional manifold and $\mathbf{g}$ a metric tensor on it, the curvature is described by the Riemann tensor $\mathbf{R}$, 
and the Einstein field equations tell us how this curvature depends on the matter sources. We can see the effects of the curvature in a space-time through the Geodesic 
Deviation Equation (GDE) \cite{Synge}-\cite{Ellis2}; this equation give us the relative acceleration of two neighbor geodesics \cite{Wald},\cite{Poisson}.
The GDE gives an elegant description of the structure of a space-time,  and all the important relations (Raychaudhuri equation, Mattig Relation, etc) could be obtained
solving the GDE for timelike, null and spacelike geodesic congruences. \\\\
Even GR is the most widely accepted gravity theory and it has been tested in several field strength regimes, is not the only relativistic theory of gravity \cite{Corda}. 
In the last decades several generalizations of the Einstein field equations have been proposed \cite{Schmidt},\cite{Wands}. Within these extended theories of gravity nowadays 
a subclass, known as $f(R)$ theories, are an alternative for classical problems, as the accelerated
expansion of the universe, instead of Dark Energy and Quintessence models \cite{Felice}-\cite{Capozziello2}. $f(R)$ theories of gravity are basically extensions of the usual Einstein-Hilbert 
action in GR with an arbitrary function of the Ricci scalar $R$ \cite{Faraoni}-\cite{Sotiriou}.  \\\\
In the context of Palatini $f(R)$ gravity the GDE had been studied in \cite{Shojai}, where they get the  modified Raychaudhuri equation. Our aim in this paper is to  obtain the GDE in the metric 
context of $f(R)$ gravity, and study some particular cases, we also obtain the generalized Mattig relation which is an useful equation to measure cosmological distances. Through this paper 
we use the sign convention $(-,+,+,+)$ and geometrical units with $c=1$.

\section{Field equations in $f(R)$ gravity}
As we mentioned above the modified theories of gravity have been studied in order to explain the
accelerated expansion of the universe, among other problems in gravity. A family of these theories is modified $f(R)$ gravity, which consists
in a generalization of the Einstein-Hilbert action (Lagrangian $\mathcal{L}[\mathbf{g}] = R-2\Lambda$), with an arbitrary function of the 
Ricci scalar ($\mathcal{L}[\mathbf{g}] = f(R)$) \cite{Nojiri},\cite{Sotiriou}.
Again, we consider the space-time as a pair $(\mathcal{M},\mathbf{g})$ with $\mathbf{g}$
a Lorentzian metric on $\mathcal{M}$; the relation of the Ricci scalar and the metric tensor is given assuming a Levi-Civita connection on the manifold. i.e. a 
Christoffel symbol. Then the action could be written with a boundary term as \cite{Guarnizo}
\begin{equation}
S = \int_{\mathcal{V}} d^4x\, \sqrt{-g}f(R)  +   2\oint_{\partial \mathcal{V}} d^3y\, \varepsilon\sqrt{|h|} f'(R)K + S_M,
\end{equation}
here $h$ is the determinant of the induced metric, $K$ is the trace of the extrinsic curvature on the boundary $\partial \mathcal{V}$, $\varepsilon$ is 
equal to $+1$ if $\partial \mathcal{V}$ is timelike and $-1$ if $\partial \mathcal{V}$ is spacelike (it is assumed that $\partial \mathcal{V}$ is nowhere null), 
and $\kappa = 8\pi G$. Finally  $S_M$ is the action for the matter fields. Variation of this action with respect to $g^{\alpha\beta}$ gives the following field 
equations \cite{Guarnizo}, which are equivalent for those without boundaries given in \cite{Buchdahl}
\begin{equation}\label{fieldequations}
\boxed{f'(R)\,R_{\alpha\beta} -\frac{f(R)}{2}\, g_{\alpha\beta} + g_{\alpha\beta}\square f'(R) - \nabla_{\alpha}\nabla_{\beta}f'(R) = \kappa T_{\alpha\beta},}
\end{equation}
where $\square  = \nabla_{\sigma}\nabla^{\sigma}$, $f'(R) = df(R)/dR$; the energy-momentum tensor is defined by
\begin{equation}\label{tensem}
T_{\alpha\beta} \equiv -2\frac{\partial \mathcal{L}_M}{\partial g^{\alpha\beta}} + \mathcal{L}_Mg_{\alpha\beta} = -\frac{2}{\sqrt{-g}}\frac{\delta S_M}{\delta g^{\alpha\beta}},
\end{equation}
being $\mathcal{L}_M$ the lagrangian for all the matter fields, we also have the conservation equation $\nabla_{\alpha}T^{\alpha\beta}=0$.  
Contracting with $g^{\alpha\beta}$ we have for the trace of the field equations
\begin{equation}\label{Trace}
f'(R)\,R -2f(R) + 3\square f'(R) = \kappa T,
\end{equation}
with  $T = g^{\alpha\beta}T_{\alpha\beta}$. From equation (\ref{fieldequations}) we can write
\begin{equation}\label{ricci}
R_{\alpha\beta}  = \frac{1}{f'(R)}\biggl[\kappa T_{\alpha\beta}+\frac{f(R)}{2}\, g_{\alpha\beta} - g_{\alpha\beta}\square f'(R) + \nabla_{\alpha}\nabla_{\beta}f'(R)\biggr],
\end{equation}
and from (\ref{Trace})
\begin{equation}\label{scalar}
R  = \frac{1}{f'(R)}\biggl[\kappa T+2f(R) - 3\square f'(R)\biggr].
\end{equation}
It is possible to write the field equations in $f(R)$ gravity, in the form of Einstein equations with an effective energy-momentum tensor \cite{Capoziello}
\begin{align}
G_{\alpha\beta} &\equiv R_{\alpha\beta} - \frac{1}{2}R\, g_{\alpha\beta}\notag\\
& = \frac{\kappa T_{\alpha\beta}}{f'(R)} + g_{\alpha\beta}\frac{[f(R) - Rf'(R)]}{2f'(R)} + \frac{[\nabla_{\alpha}\nabla_{\beta}f'(R)-g_{\alpha\beta}\square f'(R)]}{f'(R)},
\end{align}
or
\begin{equation}
G_{\alpha\beta} = \frac{\kappa}{f'(R)}\bigl(T_{\alpha\beta} + T_{\alpha\beta}^{eff}\bigr),
\end{equation}
with
\begin{equation}
T_{\alpha\beta}^{eff} \equiv \frac{1}{\kappa}\biggl[\frac{[f(R) - Rf'(R)]}{2}g_{\alpha\beta} + [\nabla_{\alpha}\nabla_{\beta}-g_{\alpha\beta}\square]f'(R)\biggr].
\end{equation}
which could be interpreted as an fluid composed by curvature terms.
\section{Geodesic Deviation Equation}
Now we give a brief discussion about the Geodesic Deviation Equation (GDE) following \cite{Ellis2}-\cite{Poisson}. Let be $\gamma_0$ and $\gamma_1$ two 
neighbor geodesics with an affine parameter $\nu$. We introduce between the two geodesics a entire family of interpolating geodesics $s$, and collectively 
describe these geodesics with $x^{\alpha}(\nu,s)$, figure 1.1. The vector field $V^{\alpha} = \frac{dx^{\alpha}}{d\nu}$ is tangent to the geodesic. 
The family $s$ has $\eta^{\alpha}= \frac{dx^{\alpha}}{ds}$ like it's tangent vector field. Thus, the acceleration for this vector field is given by \cite{Wald},\cite{Poisson}
\begin{center}
\includegraphics[scale=0.48]{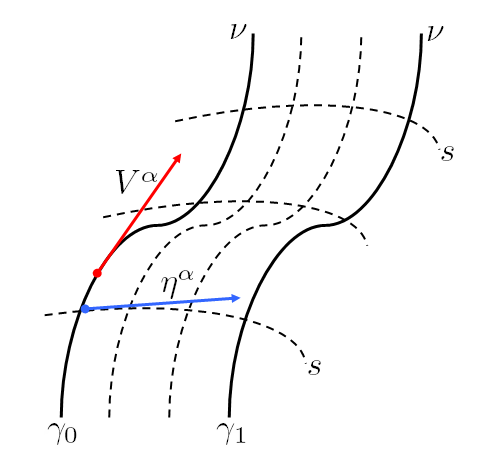}\\
{\footnotesize \textbf{Figure 1.1} Geodesic Deviation.}
\end{center}
Thus, the acceleration for this vector field is given by \cite{Wald},\cite{Poisson}
\begin{equation}\label{GDE}
\boxed{\frac{D^2 \eta^{\alpha}}{D \nu^2} = - R_{\beta\gamma\delta}^{\alpha}V^{\beta}\eta^{\gamma}V^{\delta},}
\end{equation}
which is the Geodesic Deviation Equation (GDE). Here $\frac{D}{D \nu}$ correspond to the covariant derivative a long the curve. We want to relate the geometrical properties of the space-time (Riemann and Ricci tensors) with
the matter fields through field equations. For this we write the Riemann tensor as \cite{Wald},\cite{Ellis}
\begin{equation}\label{RiemannC}
R_{\alpha\beta\gamma\delta} = C_{\alpha\beta\gamma\delta} + \frac{1}{2}\bigl(g_{\alpha\gamma}R_{\delta\beta} - g_{\alpha\delta}R_{\gamma\beta} + g_{\beta\delta}R_{\gamma\alpha}
- g_{\beta\gamma}R_{\delta\alpha} \bigr) - \frac{R}{6}\bigl(g_{\alpha\gamma}g_{\delta\beta} - g_{\alpha\delta}g_{\gamma\beta}\bigr),
\end{equation}
with $C_{\alpha\beta\gamma\delta}$ the Weyl tensor. In the case of standard cosmology (Friedman-Lama\^{i}tre-Robertson-Walker universe, hereafter FLRW universe) we have the line element
\begin{equation}
ds^2 = -dt^2 + a^2(t)\biggl[\frac{dr^2}{1-kr^2} + r^2d\theta^2 + r^2\sin^2\theta d\varphi^2\biggr],
\end{equation}
where $a(t)$ is the scale factor and $k$ the spatial curvature of the universe. In this case the Weyl tensor $C_{\alpha\beta\gamma\delta}$ vanishes, and for the energy momentum tensor we have
\begin{equation}\label{tensemfrw}
T_{\alpha\beta}  = (\rho + p)u_{\alpha}u_{\beta} + p g_{\alpha\beta},
\end{equation}
being $\rho$ the energy density and $p$ the pressure, the trace is
\begin{equation}
T = 3p-\rho.
\end{equation}
The standard form of the Einstein field equations in GR (with cosmological constant) is
\begin{equation}
\boxed{R_{\alpha\beta} - \frac{1}{2}R g_{\alpha\beta} + \Lambda g_{\alpha\beta} = \kappa T_{\alpha\beta},}
\end{equation}
then we can write the Ricci scalar $R$ and the Ricci tensor $R_{\alpha\beta}$ using (\ref{tensemfrw})
\begin{equation}
R = \kappa(\rho - 3p) + 4\Lambda,
\end{equation}
\begin{equation}
R_{\alpha\beta} = \kappa(\rho + p)u_{\alpha}u_{\beta} + \frac{1}{2}\bigl[\kappa(\rho-p) + 2\Lambda\bigr]g_{\alpha\beta},
\end{equation}
from these expressions the right side of equation (\ref{GDE}) is written as \cite{Ellis2}
\begin{equation}\label{Pirani1}
R_{\beta\gamma\delta}^{\alpha}V^{\beta}\eta^{\gamma}V^{\delta} = \biggl[\frac{1}{3}(\kappa \rho + \Lambda)\epsilon + \frac{1}{2}\kappa(\rho + p)E^2\biggr]\eta^{\alpha},
\end{equation}
with $\epsilon = V^{\alpha}V_{\alpha}$ and $E = - V_{\alpha}u^{\alpha}$. This equation is known as \textit{Pirani equation} \cite{Pirani}.
The GDE and some solutions for spacelike, timelike and null congruences has been studied in detail in
\cite{Ellis2}, which gives some important result concerning cosmological distances also showed in \cite{Ellis}.
Our purpose here is to extend these results from the modified field equations in metric $f(R)$ gravity.
\section{Geodesic Deviation Equation in $f(R)$ Gravity}
The starting point are the expressions for the Ricci tensor and the Ricci scalar from the field equations in $f(R)$ gravity, equations (\ref{ricci}) and (\ref{scalar}) respectively. 
Using the expression (\ref{RiemannC}) we can write
\begin{multline}
R_{\alpha\beta\gamma\delta} = C_{\alpha\beta\gamma\delta} + \frac{1}{2f'(R)}\Biggl[\kappa(T_{\delta\beta}g_{\alpha\gamma}-T_{\gamma\beta}g_{\alpha\delta} + T_{\gamma\alpha}g_{\beta\delta}-T_{\delta\alpha}g_{\beta\gamma}) + f(R)\bigl(g_{\alpha\gamma}g_{\delta\beta} - g_{\alpha \delta}g_{\gamma\beta}\bigr)\\
+  \bigl(g_{\alpha\gamma}\mathcal{D}_{\delta\beta}
 - g_{\alpha\delta}\mathcal{D}_{\gamma\beta} + g_{\beta\delta}\mathcal{D}_{\gamma\alpha}- g_{\beta\gamma}\mathcal{D}_{\delta\alpha}\bigr)f'(R)\Biggr] - \frac{1}{6f'(R)}\biggl(\kappa T + 2f(R) + 3\square f'(R)\biggr)\bigl(g_{\alpha\gamma}g_{\delta\beta} - g_{\alpha \delta}g_{\gamma\beta}\bigr),
\end{multline}
where we defined the operator
\begin{equation}
\boxed{\mathcal{D}_{\alpha\beta} \equiv \nabla_{\alpha}\nabla_{\beta} - g_{\alpha\beta}\square.}
\end{equation}
Thus rising the first index in the Riemann tensor and contracting with $V^{\beta}\eta^{\gamma}V^{\delta}$, the right side of the GDE could be written as
\begin{multline}\label{GeneralGDE}
R_{\beta\gamma\delta}^{\alpha}V^{\beta}\eta^{\gamma}V^{\delta} = C_{\beta\gamma\delta}^{\alpha}V^{\beta}\eta^{\gamma}V^{\delta} + \frac{1}{2f'(R)}\Biggl[\kappa(T_{\delta\beta}\delta_{\gamma}^{\alpha}-T_{\gamma\beta}\delta_{\delta}^{\alpha} + T_{\gamma}^{\, \, \alpha}g_{\beta\delta}-T_{\delta}^{\, \, \alpha}g_{\beta\gamma}) + f(R)\bigl(\delta_{\gamma}^{\alpha}g_{\delta\beta} - \delta_{\delta}^{\alpha}g_{\gamma\beta}\bigr)\\
+  \bigl(\delta_{\gamma}^{\alpha}\mathcal{D}_{\delta\beta}
 - \delta_{\delta}^{\alpha}\mathcal{D}_{\gamma\beta} + g_{\beta\delta}\mathcal{D}_{\gamma}^{\, \, \alpha}- g_{\beta\gamma}\mathcal{D}_{\delta}^{\, \, \alpha}\bigr)f'(R)\Biggr]V^{\beta}\eta^{\gamma}V^{\delta}\\ - \frac{1}{6f'(R)}\biggl(\kappa T + 2f(R) + 3\square f'(R)\biggr)\bigl(\delta_{\gamma}^{\alpha}g_{\delta\beta} - \delta_{\delta}^{\alpha}g_{\gamma\beta}\bigr)V^{\beta}\eta^{\gamma}V^{\delta},
\end{multline}
In the following section we explicitly show the steps in order to find the GDE in $f(R)$ gravity using FLRW metric, our purpose is compare with the
results from GR in the limit case $f(R) = R - 2\Lambda$.
\subsection{Geodesic Deviation Equation for the FLRW universe}
Using the FLRW metric as background we have
\begin{equation}\label{RicciRWF}
R_{\alpha\beta}  = \frac{1}{f'(R)}\biggl[\kappa(\rho + p)u_{\alpha}u_{\beta}+ \biggl(\kappa p + \frac{f(R)}{2}\biggr)g_{\alpha\beta} +
\mathcal{D}_{\alpha\beta}f'(R)\biggr],
\end{equation}
\begin{equation}\label{ScalarRWF}
R  = \frac{1}{f'(R)}\biggl[\kappa (3p-\rho) + 2f(R) - 3\square f'(R)\biggr],
\end{equation}
with these expressions the Riemann tensor could be written as
\begin{multline}
R_{\alpha\beta\gamma\delta} = \frac{1}{2f'(R)}\Biggl[\kappa(\rho+p)\bigl(u_{\delta}u_{\beta}g_{\alpha\gamma}-u_{\gamma}u_{\beta}g_{\alpha\delta} + u_{\gamma}u_{\alpha}g_{\beta\delta}-u_{\delta}u_{\alpha}g_{\beta\gamma}\bigr)\\
+ \biggl(\kappa p + \frac{\kappa\rho}{3} + \frac{f(R)}{3}+ \square f'(R)\biggr)\bigl(g_{\alpha\gamma}g_{\delta\beta} - g_{\alpha \delta}g_{\gamma\beta}\bigr)+ (g_{\alpha\gamma}\mathcal{D}_{\delta\beta}
 - g_{\alpha\delta}\mathcal{D}_{\gamma\beta} + g_{\beta\delta}\mathcal{D}_{\gamma\alpha}- g_{\beta\gamma}\mathcal{D}_{\delta\alpha})f'(R)\Biggr],
\end{multline}
If the vector field $V^{\alpha}$ is normalized, it implies $V^{\alpha}V_{\alpha} = \epsilon$, and
\begin{multline}
R_{\alpha\beta\gamma\delta}V^{\beta}V^{\delta} = \frac{1}{2f'(R)}\Biggl[\kappa(\rho+p)\bigl(g_{\alpha\gamma}(u_{\beta}V^{\beta})^2-2(u_{\beta}V^{\beta})V_{(\alpha}u_{\gamma)}
 + \epsilon u_{\alpha}u_{\gamma}\bigr)\\ + \biggl(\kappa p + \frac{\kappa\rho}{3} + \frac{f(R)}{3}+ \square f'(R)\biggr)\bigl(\epsilon g_{\alpha\gamma}-V_{\alpha}V_{\gamma}\bigr)
 + (g_{\alpha\gamma}\mathcal{D}_{\delta\beta} - g_{\alpha\delta}\mathcal{D}_{\gamma\beta} + g_{\beta\delta}\mathcal{D}_{\gamma\alpha}- g_{\beta\gamma}\mathcal{D}_{\delta\alpha})f'(R)V^{\beta}V^{\delta}\Biggr],
\end{multline}
rising the first index in the Riemann tensor and contracting with $\eta^{\gamma}$
\begin{multline}
R_{\beta\gamma\delta}^{\alpha}V^{\beta}\eta^{\gamma}V^{\delta} = \frac{1}{2f'(R)}\Biggr[\kappa(\rho+p)\bigl((u_{\beta}V^{\beta})^2\eta^{\alpha}-(u_{\beta}V^{\beta})V^{\alpha}(u_{\gamma}\eta^{\gamma})-(u_{\beta}V^{\beta})u^{\alpha}(V_{\gamma}\eta^{\gamma}) + \epsilon u^{\alpha}u_{\gamma}\eta^{\gamma}\bigr)\\
+ \biggl(\kappa p + \frac{\kappa\rho}{3} + \frac{f(R)}{3}+ \square f'(R)\biggr)\bigl(\epsilon \eta^{\alpha}-V^{\alpha}(V_{\gamma}\eta^{\gamma})\bigr) + \bigl[(\delta_{\gamma}^{\alpha}\mathcal{D}_{\delta\beta} - \delta_{\delta}^{\alpha}\mathcal{D}_{\gamma\beta} +
g_{\beta\delta}\mathcal{D}_{\gamma}^{\, \, \alpha}- g_{\beta\gamma}\mathcal{D}_{\delta}^{\, \,  \alpha})f'(R)\bigr]V^{\beta}V^{\delta}\eta^{\gamma}\Biggr],
\end{multline}
with $E = - V_{\alpha}u^{\alpha}, \eta_{\alpha}u^{\alpha}=\eta_{\alpha}V^{\alpha}=0$ \cite{Ellis2} it reduces to
\begin{multline}\label{Riemann1}
R_{\beta\gamma\delta}^{\alpha}V^{\beta}\eta^{\gamma}V^{\delta} = \frac{1}{2f'(R)}\Biggr[\kappa(\rho+p)E^2   + \epsilon\biggl(\kappa p + \frac{\kappa\rho}{3} + \frac{f(R)}{3}+ \square f'(R)\biggr)\Biggr]\eta^{\alpha}\\
+ \frac{1}{2f'(R)}\biggl[\bigl[(\delta_{\gamma}^{\alpha}\mathcal{D}_{\delta\beta} - \delta_{\delta}^{\alpha}\mathcal{D}_{\gamma\beta} +
g_{\beta\delta}\mathcal{D}_{\gamma}^{\, \, \alpha}- g_{\beta\gamma}\mathcal{D}_{\delta}^{\, \,  \alpha})f'(R)\bigr]V^{\beta}V^{\delta}\biggr]\eta^{\gamma}.
\end{multline}
From the FLRW metric the expression for the Ricci scalar is
\begin{equation}\label{RWscalar}
R = 6\biggl[\frac{\ddot{a}}{a}+ \biggl(\frac{\dot{a}}{a}\biggr)^2 + \frac{k}{a^2}\biggr] = 6\biggl[\dot{H} + 2H^2 + \frac{k}{a^2}\biggr],
\end{equation}
where we have used the definition for the Hubble parameter $H \equiv \frac{\dot{a}}{a}$. We see that $R$ is only a function of time, and only time derivatives in the operators $\mathcal{D}_{\alpha\beta}$ will not vanishing.
The non-vanishing operators are
\begin{align}\label{square}
\square f'(R) &=  - \partial_{0}^2 f'(R) - 3H \partial_{0} f'(R),\notag \\
 &=  - f''(R) \ddot{R} - f'''(R) \dot{R}^2 - 3H f''(R) \dot{R},
\end{align}
\begin{align}\label{D00}
\mathcal{D}_{00} &= -3 H \partial_{0} f'(R),\notag \\
& = -3 H f''(R) \dot{R},
\end{align}
\begin{align}
\mathcal{D}_{ij} &= 2 H g_{ij}\partial_{0} f'(R) + g_{ij}\partial_{0}^2 f'(R), \notag \\
& = 2 H g_{ij}f''(R) \dot{R} + g_{ij}(f''(R) \ddot{R} + f'''(R) \dot{R}^2),
\end{align}
being $g_{ij}$ the spatial components of the FLRW metric and $\dot{R} = \partial_{0} R$. 
With these results is easy to show that the the total contribution of the operators in (\ref{Riemann1}) is
\begin{equation}
\bigl(\delta_{\gamma}^{\alpha}\mathcal{D}_{\delta\beta} - \delta_{\delta}^{\alpha}\mathcal{D}_{\gamma\beta} + g_{\beta\delta}\mathcal{D}_{\gamma}^{\, \, \alpha}
- g_{\beta\gamma}\mathcal{D}_{\delta}^{\, \,  \alpha}\bigr)f'(R)V^{\beta}V^{\delta}\eta^{\gamma} = \epsilon \bigl(5H f''(R) \dot{R}  + f''(R) \ddot{R} + f'''(R) \dot{R}^2\bigr) \eta^{\alpha},
\end{equation}
then the expression for $R_{\beta\gamma\delta}^{\alpha}V^{\beta}\eta^{\gamma}V^{\delta}$ reduces to
\begin{equation}\label{Pirani2}
\boxed{R_{\beta\gamma\delta}^{\alpha}V^{\beta}\eta^{\gamma}V^{\delta} = \frac{1}{2f'(R)}\Biggl[\kappa(\rho+p)E^2   + \epsilon\biggl(\kappa p + \frac{\kappa\rho}{3} + \frac{f(R)}{3} + 2 H  f''(R) \dot{R}\biggr)\Biggr]\eta^{\alpha},}
\end{equation}
which is the generalization of the Pirani equation. In the particular case $f(R) = R -2\Lambda$ the previous equation reduces to (\ref{Pirani1}). \\\\
The GDE in $f(R)$ gravity from equation (\ref{GDE}) is
\begin{equation}\label{GDeFR}
\boxed{\frac{D^2 \eta^{\alpha}}{D \nu^2} = - \frac{1}{2f'(R)}\Biggl[\kappa(\rho+p)E^2   + \epsilon\biggl(\kappa p + \frac{\kappa\rho}{3} + \frac{f(R)}{3} + 2 H  f''(R) \dot{R}\biggr)\Biggr]\eta^{\alpha},}
\end{equation}
as we expect the GDE induces only a change in the magnitude of the deviation vector $\eta^{\alpha}$, which also occurs in GR.
This result is expected from the form of the metric, which describes
an homogeneous and isotropic universe. For anisotropic universes, like Bianchi I, 
the GDE also induces a \textit{change in the direction} of the deviation vector, as shown in \cite{Caceres}.
\subsection{GDE for fundamental observers}
In this case we have $V^{\alpha}$ as the four-velocity of the fluid $u^{\alpha}$. The affine parameter $\nu$ matches with the proper time of the fundamental observer
$\nu = t$. Because we have temporal geodesics then $\epsilon=-1$ and also the vector field are normalized $E = 1$ , thus from (\ref{Pirani2})
\begin{equation}\label{GDEfR1}
\boxed{R_{\beta\gamma\delta}^{\alpha}u^{\beta}\eta^{\gamma}u^{\delta} = \frac{1}{2f'(R)}\biggl[\frac{2\kappa\rho}{3}  -\frac{f(R)}{3} - 2 H  f''(R) \dot{R}\biggr]\eta^{\alpha},}
\end{equation}
if the deviation vector is $\eta_{\alpha} = \ell e_{\alpha}$, isotropy implies
\begin{equation}
\frac{D e^{\alpha}}{D t} = 0,
\end{equation}
and
\begin{equation}
\frac{D^2 \eta^{\alpha}}{D t^2} = \frac{d^2\ell}{dt^2} e^{\alpha},
\end{equation}
using this result in the GDE (\ref{GDE}) with (\ref{GDEfR1}) gives
\begin{equation}
\frac{d^2\ell}{dt^2} = - \frac{1}{2f'(R)}\biggl[\frac{2\kappa\rho}{3}  -\frac{f(R)}{3} - 2 H  f''(R) \dot{R}\biggr]\, \ell .
\end{equation}
In particular with $\ell = a(t)$ we have
\begin{equation}\label{Raycha}
\boxed{\frac{\ddot{a}}{a} = \frac{1}{f'(R)}\biggl[\frac{f(R)}{6}  + H f''(R) \dot{R} -\frac{\kappa\rho}{3} \biggr].}
\end{equation}
This equation could be obtained as a particular case of the generalized Raychaudhuri equation given in \cite{Rippl}. 
Is possible to obtain the standard form of the modified Friedmann equations \cite{Sotiriou} from this Raychaudhuri equation giving
\begin{equation}\label{ModFried1}
H^2 + \frac{k}{a^2}  = \frac{1}{3f'(R)}\biggl[\kappa\rho + \frac{(R f'(R) - f(R))}{2} - 3H f''(R) \dot{R} \biggr],
\end{equation}
and
\begin{equation}\label{ModFried2}
2\dot{H} + 3H^2 + \frac{k}{a^2}  = -\frac{1}{f'(R)}\biggl[\kappa p + 2H f''(R) \dot{R} + \frac{(f(R)- R f'(R))}{2}+ f''(R) \ddot{R} + f'''(R) \dot{R}^2\biggr].
\end{equation}

\subsection{GDE for nulll vector fields}
Now we consider the GDE for null vector fields past directed. In this case we have $V^{\alpha}=k^{\alpha}$, $k_{\alpha}k^{\alpha}=0$, then equation (\ref{Pirani2})
reduces to
\begin{equation}\label{RicciFoc}
\boxed{R_{\beta\gamma\delta}^{\alpha}k^{\beta}\eta^{\gamma}k^{\delta} = \frac{1}{2f'(R)}\kappa(\rho+p)E^2\,\eta^{\alpha},}
\end{equation}
that can be interpreted as the \textit{Ricci focusing} in $f(R)$ gravity. Writing $\eta^{\alpha}= \eta e^{\alpha}$,  $e_{\alpha}e^{\alpha}=1$, $e_{\alpha}u^{\alpha}=e_{\alpha}k^{\alpha}=0$ and choosing an aligned base parallel propagated
$\frac{D e^{\alpha}}{D \nu}=k^{\beta}\nabla_{\beta}e^{\alpha}=0$, the GDE (\ref{GDeFR}) reduces to
\begin{equation}\label{GDE3}
\frac{d^2\eta}{d\nu^2} = - \frac{1}{2f'(R)}\kappa(\rho+p)E^2\, \eta.
\end{equation}
In the case of GR discussed in \cite{Ellis2}, all families of past-directed null geodesics experience focusing, provided $\kappa(\rho+p) > 0$,
and for a fluid with equation of state $p = - \rho$ (cosmological constant) there is no influence in the focusing \cite{Ellis2}. From (\ref{GDE3}) the focusing condition for $f(R)$ gravity is
\begin{equation}
\frac{\kappa(\rho + p)}{f'(R)} > 0.
\end{equation}
A similar condition over the function $f(R)$ was established in order to avoid the appearance of ghosts \cite{Felice},\cite{Starobinsky}, in which $f'(R) > 0$. 
We want to write the equation (\ref{GDE3}) in function of the redshift parameter $z$. For this the differential operators is
\begin{equation}
\frac{d}{d\nu} = \frac{dz}{d\nu}\frac{d}{dz},
\end{equation}
\begin{align}
\frac{d^2}{d\nu^2} &= \frac{dz}{d\nu}\frac{d}{dz}\biggl(\frac{d}{d\nu}\biggr), \notag \\
& = \biggl(\frac{d\nu}{dz}\biggr)^{-2}\biggl[-\biggl(\frac{d\nu}{dz}\biggr)^{-1}\frac{d^2\nu}{dz^2}\frac{d}{dz}+\frac{d^2}{dz^2}\biggr].
\end{align}
In the case of null geodesics
\begin{equation}
(1+z) = \frac{a_0}{a}=\frac{E}{E_0} \quad \longrightarrow \quad \frac{dz}{1+z}=- \frac{da}{a},
\end{equation}
with $a$ the scale factor, and $a_0=1$ the present value of the scale factor. Thus for the past directed case
\begin{equation}
dz = (1+z) \frac{1}{a}\frac{da}{d\nu}\, d\nu = (1+z) \frac{\dot{a}}{a} E\, d\nu = E_0 H (1+z)^2\, d\nu,
\end{equation}
Then we get
\begin{equation}
\frac{d\nu}{dz} = \frac{1}{E_0 H (1+z)^2},
\end{equation}
and
\begin{equation}
\frac{d^2\nu}{dz^2} =- \frac{1}{E_0 H (1+z)^3}\biggl[\frac{1}{H}(1+z)\frac{dH}{dz}+2\biggr],
\end{equation}
writing $\frac{dH}{dz}$ as
\begin{equation}
\frac{dH}{dz} = \frac{d\nu}{dz}\frac{dt}{d\nu}\frac{dH}{dt} = - \frac{1}{H(1+z)} \frac{dH}{dt},
\end{equation}
the minus sign comes from the condition of past directed geodesic, when $z$ increases, $\nu$ decreases. We also use $\frac{dt}{d\nu} = E_0 (1+z)$.
Now, from the definition of the Hubble parameter $H$
\begin{equation}
\dot{H} \equiv \frac{dH}{dt} = \frac{d}{dt}\frac{\dot{a}}{a} = \frac{\ddot{a}}{a} - H^2,
\end{equation}
and using the Raychaudhuri equation (\ref{Raycha})
\begin{equation}
\dot{H} = \frac{1}{f'(R)}\biggl[\frac{f(R)}{6}  + H f''(R) \dot{R} -\frac{\kappa\rho}{3} \biggr]- H^2,
\end{equation}
then
\begin{equation}
\frac{d^2\nu}{dz^2} = -\frac{3}{E_0 H (1+z)^3}\biggl[1+ \frac{1}{3H^2 f'(R)}\biggl(\frac{\kappa \rho}{3}- \frac{f(R)}{6} - H f''(R) \dot{R}\biggr)\biggr].
\end{equation}
Finally, the operator $\frac{d^2\eta}{d\nu^2}$ is
\begin{equation}
\frac{d^2\eta}{d\nu^2} = \bigl(EH(1+z)\bigr)^2\Biggl[\frac{d^2\eta}{dz^2} + \frac{3}{(1+z)}\biggl[1+ \frac{1}{3H^2 f'(R)}\biggl
(\frac{\kappa \rho}{3}- \frac{f(R)}{6} - H f''(R) \dot{R}\biggr)\biggr]\frac{d\eta}{dz}\Biggr],
\end{equation}
and the GDE (\ref{GDE3}) reduces to
\begin{equation}
\boxed{\frac{d^2\eta}{dz^2} + \frac{3}{(1+z)}\Biggl[1+ \frac{1}{3H^2 f'(R)}\biggl
(\frac{\kappa \rho}{3}- \frac{f(R)}{6} - H f''(R) \dot{R}\biggr)\Biggr]\, \frac{d\eta}{dz} + \frac{\kappa(\rho+p)}{2H^2(1+z)^2f'(R)}\, \eta = 0,}
\end{equation}
The energy density $\rho$ and the pressure $p$ considering the contributions from matter an radiation could be written in the following way
\begin{equation}
\kappa\rho = 3H_0^2 \Omega_{m0}(1 + z)^3 + 3H_0^2\Omega_{r0}(1 + z)^4, \qquad \kappa p = H_0^2\Omega_{r0}(1 + z)^4,
\end{equation}
where we have used $p_m=0$ and $p_r = \frac{1}{3}\rho_r$. Thus the GDE could be written as
\begin{equation}\label{MattigGen}
\frac{d^2\eta}{dz^2} + \mathcal{P}(H,R,z)\frac{d\eta}{dz} + \mathcal{Q}(H,R,z)\eta = 0,
\end{equation}
with
\begin{equation}
\boxed{\mathcal{P}(H,R,z) = \frac{4\Omega_{m0}(1 + z)^3 + 4\Omega_{r0}(1 + z)^4 +  3f'(R)\Omega_{k0}(1 + z)^2  + 4\Omega_{DE} -\frac{Rf'(R)}{6 H_0^2}}{(1+z)\bigl(\Omega_{m0}(1 + z)^3 + \Omega_{r0}(1 + z)^4 + f'(R)\Omega_{k0}(1 + z)^2 + \Omega_{DE}\bigr)},}
\end{equation}
\begin{equation}
\boxed{\mathcal{Q}(H,R,z) = \frac{3\Omega_{m0}(1 + z) + 4\Omega_{r0}(1 + z)^2}{2\bigl(\Omega_{m0}(1 + z)^3+\Omega_{r0}(1 + z)^4 +  f'(R)\Omega_{k0}(1 + z)^2 + \Omega_{DE}\bigr)}.}
\end{equation}
and $H$ given by the modified field equations (\ref{ModFried1})
\begin{align}\label{Friedmod}
H^2 &= \frac{1}{f'(R)}\biggl[H_0^2 \Omega_{m0}(1 + z)^3 + H_0^2\Omega_{r0}(1 + z)^4 + \frac{(R f'(R)-f(R))}{6} - H f''(R) \dot{R}\biggr] - \frac{k}{a^2}, \notag \\
H^2 &= H_0^2\biggl[\frac{1}{f'(R)}\bigl(\Omega_{m0}(1 + z)^3 + \Omega_{r0}(1 + z)^4 + \Omega_{DE} \bigr) + \Omega_{k}(1+z)^2\biggr],
\end{align}
where
\begin{equation}\label{OmegaDE}
\boxed{\Omega_{DE} \equiv  \frac{1}{H_0^2 }\biggl[\frac{(R f'(R)-f(R))}{6} - H f''(R) \dot{R}\biggr],}
\end{equation}
and
\begin{equation}
\Omega_{k0}=-\frac{k}{H_0^2 a_0^2}.
\end{equation}
In order to solve (\ref{MattigGen}) it is necessary to write $R$ and $H$ in function of the redshift.
First we define the operator
\begin{equation}
\frac{d}{dt} = \frac{dz}{da}\frac{da}{dt}\frac{d}{dz} = -(1+z)H \frac{d}{dz},
\end{equation}
then the Ricci is \cite{Capozziello2}
\begin{align*}
R &= 6\biggl[\frac{\ddot{a}}{a}+ \biggl(\frac{\dot{a}}{a}\biggr)^2 + \frac{k}{a^2}\biggr], \notag \\
&=6\biggl[2H^2 + \dot{H} + \frac{k}{a^2}\biggr], \notag \\
&=6\biggl[2H^2  -(1+z)H \frac{dH}{dz} + k(1+z)^2 \biggr],
\end{align*}
if we want $H = H(z)$ is necessary to fix the form of $H(z)$ or either a specific form of the $f(R)$ function. This point has been studied in \cite{Capozziello2} 
and the method to fix the form of $H(z)$ and find the form of the $f(R)$ function by observations is given in \cite{Capoziello3}.\\\\
In the particular case $f(R) = R -2 \Lambda$, implies $f'(R) = 1$, $f''(R) =0$. The expression for $\Omega_{DE}$ reduces to
\begin{equation}
\Omega_{DE} = \frac{1}{H_0^2 }\biggl[\frac{(R-R + 2\Lambda)}{6}\biggr] = \frac{\Lambda}{3H_0^2} \equiv \Omega_{\Lambda},
\end{equation}
then the quantity $\Omega_{DE}$ generalizes the Dark Energy parameter. The Friedmann modified equation (\ref{Friedmod}) reduces to the well know expression in GR
\begin{equation}
H^2 = H_0^2\bigl[\Omega_{m0}(1 + z)^3 + \Omega_{r0}(1 + z)^4 + \Omega_{\Lambda}  + \Omega_{k}(1+z)^2\bigr],
\end{equation}
the expressions $\mathcal{P}$, and $\mathcal{Q}$ reduces to
\begin{equation}
\mathcal{P}(z) = \frac{4\Omega_{r0}(1 + z)^4 + (7/2)
\Omega_{m0}(1 + z)^3 + 3\Omega_{k0}(1 + z)^2 + 2\Omega_{\Lambda}
}{(1+z)\bigl(\Omega_{r0}(1 + z)^4 + \Omega_{m0}(1 + z)^3 + \Omega_{k0}(1 + z)^2 + \Omega_{\Lambda}\bigr)},
\end{equation}
\begin{equation}
\mathcal{Q}(z) = \frac{2\Omega_{r0}(1 + z)^2 + (3/2)
\Omega_{m0}(1 + z)}{\Omega_{r0}(1 + z)^4 + \Omega_{m0}(1 + z)^3 + \Omega_{k0}(1 + z)^2 + \Omega_{\Lambda}}.
\end{equation}
and the GDE for null vector fields is
\begin{multline}
\frac{d^2\eta}{dz^2} + \frac{4\Omega_{r0}(1 + z)^4 + (7/2)
\Omega_{m0}(1 + z)^3 + 3\Omega_{k0}(1 + z)^2 + 2\Omega_{\Lambda}
}{(1+z)\bigl(\Omega_{r0}(1 + z)^4 + \Omega_{m0}(1 + z)^3 + \Omega_{k0}(1 + z)^2 + \Omega_{\Lambda}\bigr)}\, \frac{d\eta}{dz}\\ + \frac{2\Omega_{r0}(1 + z)^2 + (3/2)
\Omega_{m0}(1 + z)}{\Omega_{r0}(1 + z)^4 + \Omega_{m0}(1 + z)^3 + \Omega_{k0}(1 + z)^2 + \Omega_{\Lambda}}\, \eta = 0.
\end{multline}
The Mattig relation in GR is obtained in the case $\Omega_{\Lambda} = 0$ and writing $\Omega_{k0}=1-\Omega_{m0} - \Omega_{r0}$ which gives \cite{Ellis2}
\begin{equation}
\frac{d^2\eta}{dz^2} + \frac{6 +
\Omega_{m0}(1 + 7z) + \Omega_{r0}(1 + 8z + 4z^2)}{2(1 + z)(1 + \Omega_{m0}z +
\Omega_{r0}z(2 + z))}\, \frac{d\eta}{dz} + \frac{3\Omega_{m0} + 4\Omega_{r0}(1 + z)}{2(1 + z)(1 + \Omega_{m0}z +
\Omega_{r0}z(2 + z))}\, \eta = 0,
\end{equation}
then, the equation (\ref{MattigGen}) give us a generalization of the Mattig relation in $f(R)$ gravity.\\\\
In a spherically symmetric space-time, like in FLRW universe, the magnitude of the deviation vector $\eta$ is related with the proper area $dA$ of a 
source in a redshift $z$ by $d\eta \propto \sqrt{dA}$, and from this, the definition of the angular diametral distance $d_A$ could be written as \cite{Schneider}
\begin{equation}
d_A = \sqrt{\frac{dA}{d\Omega}},
\end{equation}
with $d\Omega$ the solid angle. Thus the GDE in terms of the angular diametral distance is
\begin{multline}\label{diametrald}
\frac{d^2\, d_{A}^{f(R)}}{dz^2} + \frac{4\Omega_{m0}(1 + z)^3 + 4\Omega_{r0}(1 + z)^4 +  3f'(R)\Omega_{k0}(1 + z)^2
+ 4\Omega_{DE} -\frac{Rf'(R)}{6 H_0^2}}{(1+z)\bigl(\Omega_{m0}(1 + z)^3 + \Omega_{r0}(1 + z)^4 + f'(R)\Omega_{k0}(1 + z)^2 + \Omega_{DE}\bigr)}\, \frac{d \, d_{A}^{f(R)}}{dz}\\ +
\frac{3\Omega_{m0}(1 + z) + 4\Omega_{r0}(1 + z)^2}{2\bigl(\Omega_{m0}(1 + z)^3+\Omega_{r0}(1 + z)^4 +  f'(R)\Omega_{k0}(1 + z)^2 + \Omega_{DE}\bigr)}\, d_{A}^{f(R)} = 0,
\end{multline}
where we denote the de angular diametral distance by $d_{A}^{f(R)}$, to emphasize that any solution of the previous 
equation needs a specific form of the $f(R)$ function, or either a form of $H(z)$.  This equation satisfies the initial conditions (for $z \geq z_0$)
\begin{equation}\label{condition1}
d_A^{f(R)}(z,z_0)\biggl|_{z=z_0} = 0,
\end{equation}
\begin{equation}\label{condition2}
\frac{d\, d_A^{f(R)}}{dz}(z,z_0)\biggl|_{z=z_0} = \frac{H_0}{H(z_0) (1+z_0)},
\end{equation}
being $H(z_0)$ the modified Friedmann equation (\ref{Friedmod}) evaluated at $z=z_0$.

\section{Dyer-Roeder like Equation in $f(R)$ Gravity}
Finally we get an important relation that is a tool to study cosmological distances also in inhomogeneous universes.
The Dyer-Roeder equation gives a differential equation for the diametral angular distance $d_A$ as a function of the redshift $z$ \cite{Dyer}. 
The standard form of the Dyer-Roeder equation in GR can be given by \cite{Castaneda},\cite{Okamura}
\begin{equation}
(1+z)^2 \mathcal{F}(z) \frac{d^2\, d_A}{dz^2} + (1+z) \mathcal{G}(z) \frac{d\, d_A}{dz^2} + \mathcal{H}(z) d_A = 0
\end{equation}
with
\begin{equation}
\mathcal{F}(z) = H^2(z)
\end{equation}
\begin{equation}
\mathcal{G}(z) = (1+z)H(z) \frac{dH}{dz} + 2 H^2(z)
\end{equation}
\begin{equation}
\mathcal{H}(z) = \frac{3 \tilde{\alpha}(z)}{2}\Omega_{m0}(1+z)^3
\end{equation}
with $\tilde{\alpha}(z)$ is the \textit{smoothness parameter}, which gives the character of inhomogeneities in the energy density \cite{Schneider}. There have been some studies about the influence 
of the smoothness parameter $\tilde{\alpha}$ in the behavior of $d_A(z)$ \cite{Schneider},\cite{Castaneda},\cite{Linder}.
In order to obtain the Dyer-Roeder like equation in $f(R)$ gravity we follow \cite{Schneider}.
First, we note that the terms containing the derivatives of $d_{A}^{f(R)}$ in equation (\ref{diametrald}) comes from the transformation $\frac{d}{d\nu} \longrightarrow \frac{d}{dz}$ and the term with only
$d_{A}^{f(R)}$ comes from the Ricci focusing (\ref{RicciFoc}). Then following \cite{Dyer},\cite{Castaneda} we introduce a mass-fraction $\tilde{\alpha}$ (smoothness parameter) of matter in the universe, and then we replace
\textit{only} in the Ricci focusing $\rho \longrightarrow \tilde{\alpha}\rho$. Thus following our arguments to obtain the equation (\ref{MattigGen}), and considering the case $\Omega_{r0}=0$ we get
\begin{multline}\label{DyerRoederMod}
(1+z)^2\frac{d^2\, d_{A}^{f(R),\text{ DR}}}{dz^2} + (1+z)\frac{4\Omega_{m0}(1 + z)^3 + 3f'(R)\Omega_{k0}(1 + z)^2
+ 4\Omega_{DE} -\frac{Rf'(R)}{6 H_0^2}}{\bigl(\Omega_{m0}(1 + z)^3 + f'(R)\Omega_{k0}(1 + z)^2 + \Omega_{DE}\bigr)}\, \frac{d \, d_{A}^{f(R), \text{ DR}}}{dz}\\ +
\frac{3\tilde{\alpha}(z)\Omega_{m0}(1 + z)^3 }{2\bigl(\Omega_{m0}(1 + z)^3 +  f'(R)\Omega_{k0}(1 + z)^2 + \Omega_{DE}\bigr)}\, d_{A}^{f(R), \text{ DR}} = 0.
\end{multline}
where we denote the Dyer-Roeder distance in $f(R)$ gravity by $d_{A}^{f(R),\text{ DR}}$. This Dyer-Roeder also satisfies the conditions (\ref{condition1}) and (\ref{condition2}), and it reduces to the standard form of GR in the case $f(R) = R - 2\Lambda$.\\\\
Is important to notice that the $f(R)$ function and it's derivatives appear explicitly in (\ref{diametrald}) and (\ref{DyerRoederMod}), 
being this a big difference respect to the case of GR. Another feature for these theories is the role of the dark energy parameter $\Omega_{DE}$, 
which contains the higher order terms of the field equations\footnote{From equation (\ref{OmegaDE}) the terms containing time derivatives of the Ricci scalar comes from the operator $\mathcal{D}_{00}$, 
equation (\ref{D00}), which gives the fourth order terms in the field equations of $f(R)$ gravity.}. 
However, in order to show the effects of the higher order terms, is necessary, as we stressed above, fix either $H(z)$ of $f(R)$. Further studies are necessary to quantify these effects.

\section{Conclusions and Discussion}
We have studied the Geodesic Deviation Equation (GDE) in the metric formalism of $f(R)$ gravity, starting for this from the form of the Ricci tensor and the Ricci scalar with the modified field equations. A generalized expression
for the force term (Pirani equation) is obtained for $f(R)$ gravity (for the FLRW universe), as well as the generalized GDE, both expressions reduce to the standard forms in the case $f(R) = R -2 \Lambda$.
Then, we consider two particular cases, the GDE for fundamental observers and the GDE for null vector field past directed. \\\\
The case of fundamental observers give us the Raychaudhuri equation and from this the standard form of the modified field equations
with the FLRW metric as background. The null vector field case give us more important results: a generalization of the Mattig relation,  the differential equation for the diametral angular distance in $f(R)$ gravity, and as extension we consider also the Dyer-Roeder like equation. \\\\
Another important result comes from the definition of the Ricci focusing: in the case of GR (in the FLRW space-time) we have a focusing for past-directed null geodesics for the condition $\kappa(\rho + p) > 0$, here the condition
is given by $\kappa(\rho + p)/f'(R) > 0$. \\\\
In the other hand, the general equation (\ref{GeneralGDE}) is the starting point for important issues in $f(R)$ gravity, as singularity theorems \cite{Hawking} and gravitational lensing \cite{Schneider}. We could study also the GDE for more general space-times (like Bianchi models) and for different forms of the energy-momentum tensor, in the context of $f(R)$ gravity.\\\\
Finally, the solution of equations (\ref{diametrald}) and (\ref{DyerRoederMod}) allows a different test for cosmological observables (areas, volumes, proper distances, etc.) in $f(R)$ gravity, being an alternative method to constrain some models through cosmological probes.

\end{document}